\documentclass[fleqn,usenatbib]{mnras}

\usepackage{newtxtext,newtxmath}
\usepackage{subfigure}

\usepackage[T1]{fontenc}

\DeclareRobustCommand{\VAN}[3]{#2}
\let\VANthebibliography\thebibliography
\def\thebibliography{\DeclareRobustCommand{\VAN}[3]{##3}\VANthebibliography}

\usepackage{graphicx}	
\usepackage{amsmath}	

\title[Action of successive buckling on bar and disk]{How do the successive buckling events affect a galaxy bar and stellar disk? : Potential Observable Signatures For Spotting the Buckling Action -I}

\author[Kataria, S. K.]{
Sandeep Kumar Kataria,$^{1,2}$\thanks{E-mail: skkataria.iit@gmail.com}
\\
$^{1}$Department of Astronomy, School of Physics and Astronomy, Shanghai Jiao Tong University, 800 Dongchuan Road, Shanghai 200240, China\\
$^{2}$Key Laboratory for Particle Astrophysics and Cosmology (MOE) / Shanghai Key Laboratory for Particle Physics and Cosmology, Shanghai 200240, China\\
$^{3}$ Department of Space, Planetary \& Astronomical Sciences and Engineering, Indian Institute of Technology Kanpur, Kanpur 208016, India\\
}

\date{Accepted XXX. Received YYY; in original form ZZZ}

\pubyear{2015}

\begin{document}
\label{firstpage}
\pagerange{\pageref{firstpage}--\pageref{lastpage}}
\maketitle

\begin{abstract} 
Until now, observations have caught up only a handful of galaxies in ongoing buckling action. Interestingly, N-body simulations over the past several decades show that almost every bar buckles or vertically thickens as soon as it reaches its peak strength during its evolution and leads to box/peanut/x (BPX) shapes. In order to understand the effect of multiple buckling events on the observable properties of galactic bar and disk, we perform an N-body simulation of a Milky Way-type disk. The axisymmetric galaxy disk forms a bar within a Gyr of its evolution and the bar undergoes two successive buckling events. We report that the time spans of these two buckling events are 220 Myr and 1 Gyr which have almost similar strengths of the bending modes. As a result of these two buckling events, the full lengths of BPX shapes are around 5.8 kpc and 8.6 kpc which are around two-thirds of the full bar length at the end of each buckling event. We find that the first buckling occurs at a smaller scale (radius $<$ 3 kpc) with a shorter time span affecting the larger length scales of the disk which is quantified in terms of changes in $m=$2 and $m=$ 4 Fourier modes.  While the second buckling occurs at larger scales (radius $\approx$ 6 kpc) affecting the inner disk the most. Finally, we provide observable kinematic signatures (i.e. quadrupolar patterns of the line-of-sight velocities) which can potentially differentiate the successive buckling events.

\end{abstract}




\section{Introduction} \label{intro}

Most of the disk galaxies in the local universe are barred ones which have been observed mainly in the optical and infrared wavelengths \citep{Eskridge.et.al.2000,Delmestre.et.al.2007, Sheth.et.al.2008, Erwin.2018, Lee.et.al.2019}. The evidence for significant bar fractions at higher redshifts \citep{Jogee.etal.2004,Sheth.et.al.2008} suggests bars can be long-lived. Recent JWST observations quantify the barred galaxies at $z>2$ \citep{Guo.et.al.2023, Le.Conte.2024} which suggests bars play a significant role in galaxy evolution. This agrees with the high bar fraction at large redshifts in the next-generation cosmological simulations \citep{Rosas-Guevara.et.al.2020, Zhao.et.al.2020, Yetli.et.al.2022, Ansar.et.al.2023_Fire, Ansar.Kataria.Das.2023, Kataria.Vivek.2024, Fragkoudi.et.al.2024}.

The bars are mostly associated with boxy/peanut/x-shaped bulges (BPX) in massive disks which is not the case for less massive disks \citep{Erwin.Debattista.2017, Li_2017,Erwin.et.al.2023}. Observations show the massive disks of the Milky Way \citep{Ness.Lang.2016} and M31 \citep{Beaton.et.al.2007, Blana.et.al.2018} possess BPX shapes associated with bars. Further, studies \citep{Erwin.Debattista.2017, Li_2017,Erwin.et.al.2023} also claim that the presence of BPX bulge is dominated in gas-poor galaxy disks which is result of constraints on the disk masses. BPX bulges are observed in around 50 $\%$  of edge-on disk galaxies \citep{Lutticke.et.al.2000}. BPX  bulges have been also detected over a range of redshifts from z=0 to z=1 \citep{Kruk.et.al.2019, Kumar.Kataria.2022} which shows decreasing frequency with redshifts.   

Over the past several decades N-body simulations have shown that the bar grows in the disk by transferring angular momentum from the inner disk to the outer disk and the surrounding dark matter halo \citep{Athanassoula.2003, Sellwood.et.al.2014}. The central massive concentration has been shown to suppress the bar instability \citep{Ostriker.Peebles.1973, Shen.Sellwood.2004, Kataria.Das.2018, Saha.elmegreen.2018, Kataria.Das.2019, Kataria.et.al.2020} while the role of halo spin on bar suppression is debated \citep{Collier.etal.2018,Kataria.Shen.2022,Xingchen.et.al.2023A, Chiba.Kataria.2024,Kataria.Shen.2024ApJ}.
 One of the most common features associated with the bar is the BPX shape \citep{Combes.etal.1990} that can result from the vertical thickening of planar bar orbits. The vertical thickening can result from three processes \citep{Sellwood.Gerhard.2020} namely; vertical buckling \citep{Raha.etal.1991}, vertical heating of 2:1 resonance \citep{Pfenniger.Friedli.1991, Quillen.etal.2014} and gradual trapping of 2:1 orbits resonances. The vertical buckling of the bar is similar to the bending modes in stellar disks \citep{Sellwood.Merritt.1994} though underlying physics is yet to be explored \citep{Xingchen.et.al.2023B}.  The buckling of the bar leads to weakening in the size and strength of the bar and results in  BPX shape \citep{Debattista.etal.2005, MArtinez-Valpuesta.Sholsman.2004}. The weaker buckling has been shown to destroy the bar completely \citep{Collier.2020}. Recent study \citep{Behzad.et.al.2024} claims that the BPX morphology also depend on the bar angular momentum.
 
 The recurrent buckling events have been reported first by \cite{Martinez-Valpuesta.etal.2006} and successively many studies reported two buckling events in disk evolution \citep{Collier.etal.2018, Lokas.2019B, Kataria.Shen.2022}. \cite{KumarA.et.al.2022} report the three successive buckling events for the first time in galaxies with prolate dark matter halos. The most common feature has been that the successive buckling tends to have larger time spans and act at larger length scales in the disk \citep{Athanassoula2016book}. Further, many studies have looked at the effect of classical bulges \citep{Smirnov.2019}, gas fractions \citep{Lokas.2020} and tidal interactions \citep{Lokas.2019A,Kumar.etal.2021} on the timescale and strengths of the buckling events. The fraction of enhancing classical bulge and gas components leads to suppressing buckling while tidal interaction enhances the timescale of buckling event.

In this work, we aim to understand the role of multiple buckling events on the galaxy bar and disk. We motivate our study with the question: ``How do the successive bending modes of bar for a given amplitude affect the  observable features of the bar and disk ?'' The buckling event transfers the disk's planar energy into vertical energy. Therefore, we are also interested to know which length scales in the disk are affected by these successive buckling modes of the bar. In order to understand this, we perform an N-body simulation of a Milky Way-type disk galaxy. The plan of this paper is as follows. Section \ref{methods} describes the details of model and simulation, section \ref{Results} contains the evolution of bar properties and disk during its evolution, section \ref{discussion} mentions the implication of this study and finally section \ref{conclusion} concludes with the key results.

\section{Simulation of N-body galaxy model} \label{methods}
We generate the Milky Way-like galaxy N-body model using GAlIC \citep{Yurin.Springel.2014} code. The galaxy model has a stellar disk and dark matter halo components with a total galaxy mass of 6.38 $\times$ $10^{11} M_{\sun}$. There are 1 million particles in each component of the galaxy model. The stellar disk mass fraction is 0.1 of the total galaxy mass. The galaxy has an exponential disk with radial disk scale length $R_d=2.9$ kpc. The Hernquist dark matter halo, with an inner profile similar to the NFW halo \citep{Springel.etal.2005}, has a virial radius of $140$ kpc. The galaxy model used in this work is taken from our earlier work \citep{Kataria.Shen.2022} which is referred to as the S075 model in the previous work. We refer the reader to previous work for more details on the initial condition of the galaxy models used here.

We ran the N-body simulation using GADGET-2 code \citep{Volker.2005} under self-gravity and evolved the model until 9.78 Gyr. The code makes use of the tree method \citep{Barnes.Hut.1986} to compute gravitational forces which optimizes the order of calculations. The tree opening angle is set to be $\theta_{tot}=$ 0.4 which has improved force calculation accuracy. The softening length for the halo and disk components have been chosen as 30 and 25 pc respectively. We have used the values of the time step parameter to be $\eta< = $~0.15 and the force accuracy parameter $< =$~0.0005 in order to improve the force calculations. The total angular momentum of the disk and halo is conserved to within 0.1 $\%$ throughout the simulation. We provide all of our results using code units in this article. The GAlIC and GADGET-2 codes have a unit mass equal to $10^{10}$  $M_{\sun}$ and the unit length is 1 kpc.

\section{Results}\label{Results}
\begin{figure}
    \centering
    \includegraphics[scale=0.52]{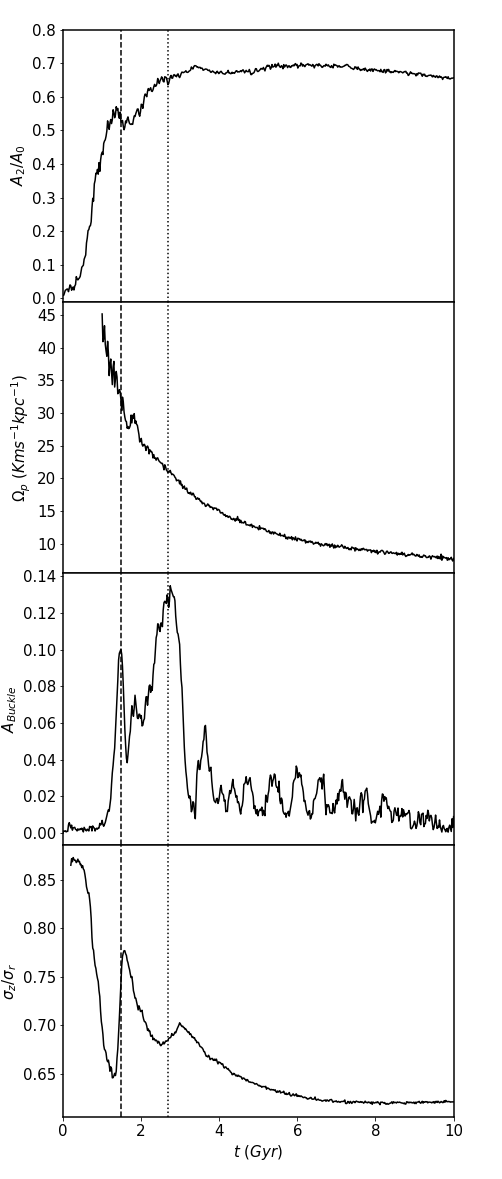}
    \caption{The time evolution of bar strength ($A_2/A_0$), pattern speed ($\Omega_p$), buckling strength ($A_{Buckle}$) and $\sigma_z/\sigma_r$ ($r<2*R_d$) is shown from top to bottom panel respectively. Here vertical dashed line corresponds to the first buckling ($\approx$ 1.45 Gyr) and the vertical dotted line corresponds to the second buckling event ($\approx$ 2.68 Gyr). }
    \label{fig:BSPSBU}
\end{figure}

\begin{figure}
    \centering
    \includegraphics[scale=0.5]{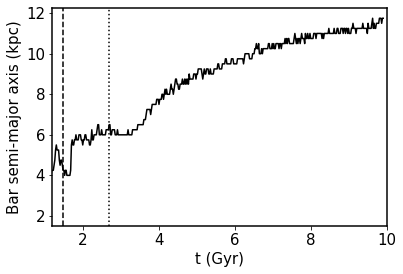}
    \caption{The time evolution of bar semi-major axis (half of the full barlength) with time has been shown here. The dashed and dotted vertical lines corresponds to first ($\approx$ 1.45 Gyr) and second buckling ($\approx$ 2.68 Gyr) events respectively.}
    \label{fig:barlength}
\end{figure}

\begin{figure*}
   \begin{tabular}{c|c}
   
    \subfigure{\includegraphics[width=0.5\textwidth]{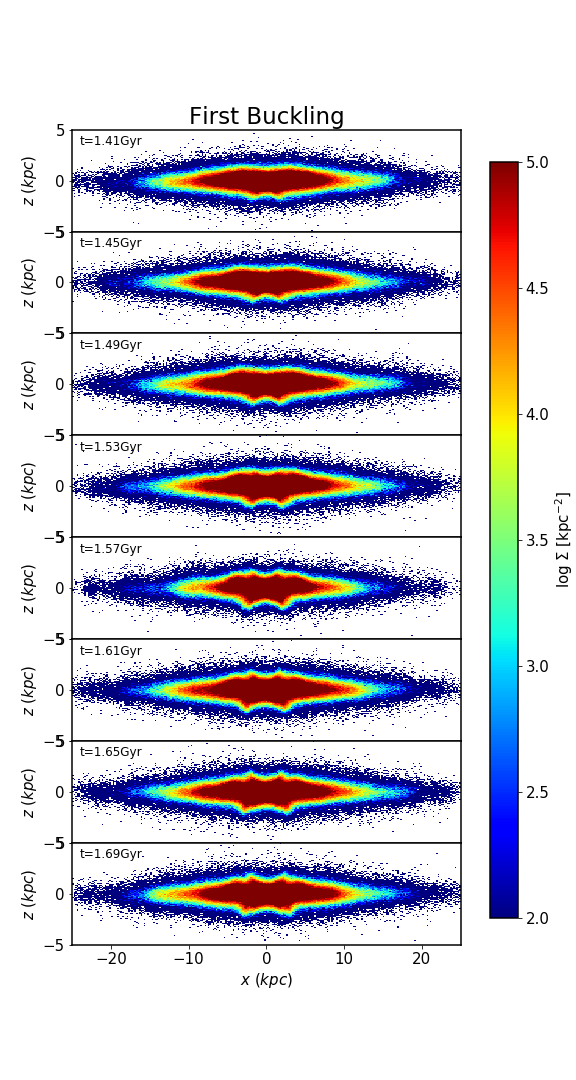}
    }   
    \subfigure{\includegraphics[width=0.5\textwidth]{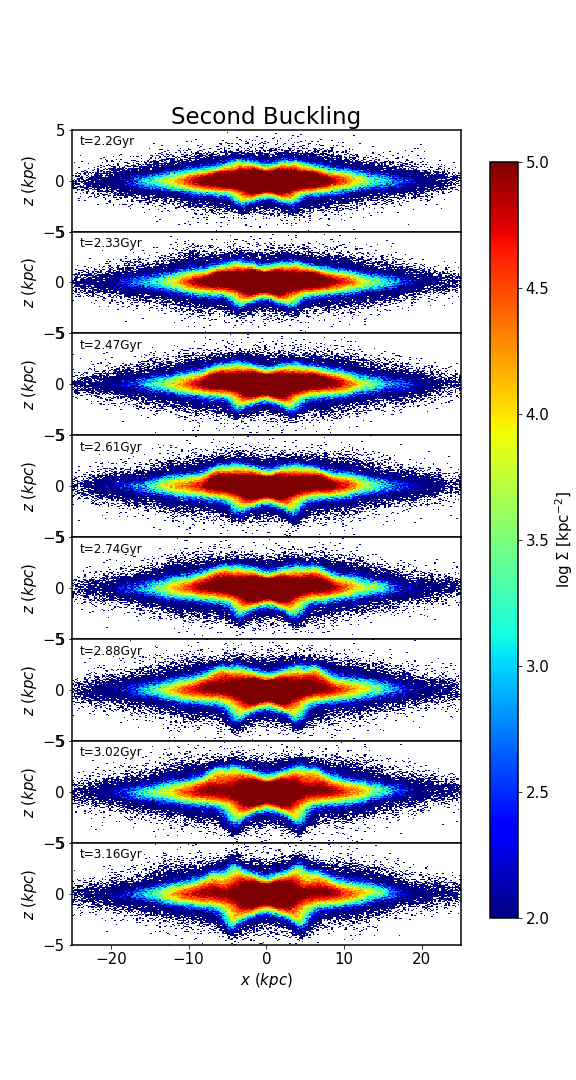}}    

   \end{tabular}

    \caption{ Left panel shows the time evolution of edge-on stellar mass density during the first buckling event which lasts for around 220 Myr. The right panel shows similar stellar mass densities for the second buckling which last for around 1 Gyr. In all of the above maps, the bar's major axis is aligned along the x-axis. }
    \label{fig:Edgeon_maps}
\end{figure*}
We find that our N-body model forms a bar within a Gyr of its evolution. In order to study the evolution of the galaxy model with time, we measure the various bar properties like bar strength, pattern speed and buckling strength in the following manner. 
 We look for $m$=2 Fourier mode in stellar disk as a proxy for bar strength which is similar to those used in literature \citep{Combes.Sanders.1981,Athanassoula.2003}.

\begin{equation}
a_2(R)=\sum_{i=1}^{N}  m_i \cos(2\phi_i)\\ \hspace{0.5cm}
b_2(R)=\sum_{i=1}^{N} m_i \sin(2 \phi_i)
\label{equation:FM}
\end{equation}

where $a_2$ and $b_2$ are calculated for the disk particles in concentric radial bins of 1 kpc throughout the disk, $m_i$ is mass of $i^{th}$ star, $\phi_i$ is the azimuthal angle. We have defined the bar strength as the maximum value of normalised $A_2$ mode among all the concentric bins of 1 kpc size. 
\begin{equation}
\frac{A_2}{A_0}= max \Bigg(\frac{\sqrt{a_2 ^2 +b_2 ^2}}{\sum_{i=1}^{N} m_i} \Bigg)
\label{eq:barstrength}
\end{equation}

We have defined the bar semi-major axis as the radius of the bin where the $\frac{A_2}{A_0}$ is equal to 60 \% of the peak value of $\frac{A_2}{A_0}$. This radius is larger than the radius corresponding to the peak value of $\frac{A_2}{A_0}$ from the centre of the galaxy. The full bar length is twice that of the bar semi-major axis.

Pattern speed ($\Omega_P$) of the bar has been measured as the change in phase angle  $\phi=\frac{1}{2}\tan^{-1}\bigg(\frac{b_2}{a_2}\bigg)$ of the bar with time where $a_2$ and $b_2$ are defined by equation \ref{equation:FM}. In order to measure the phase angle of the bar we have used the Fourier mode of the annular region corresponding to the peak in maximum value of $A_2/A_0$. 

 In order to measure the vertical buckling or bending of the bar, we have used the buckling amplitude which is given by the following equation \citep{Debattista.etal.2006}.
 \begin{equation}
    A_{Buckle}= \left | \dfrac{\sum_{i=1}^{N} z_im_i e^{2i\phi_{i}} }{\sum_{i=1}^{N} m_i} \right |
    \label{eq:Buckle}
\end{equation}
Where z is the vertical coordinate, $m_i$ is the mass of $i^{th}$ star, $\phi_i$ is the azimuthal angle.
\begin{figure*}
    \centering
    \includegraphics[scale=0.5]{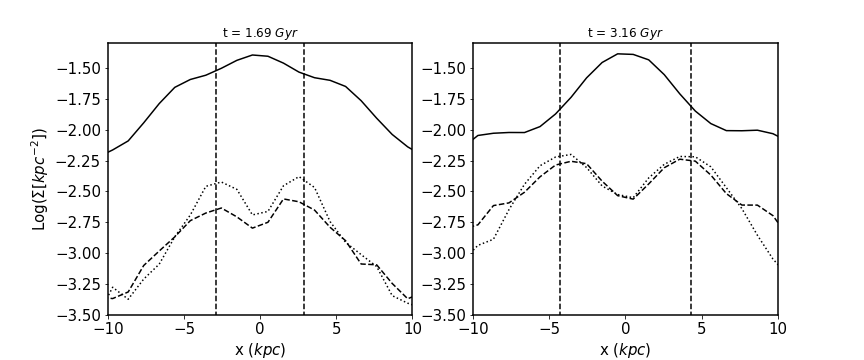}
    \caption{The line density profiles along $z=$0 (solid line), $z=$1.5 kpc (dashed line), $z=$-1.5 (dotted line) just after the two buckling events in x-z plane of the disk. The vertical dashed line corresponds to peaks in the line profile above and below mid plane of the disk. The region between these vertical dashed lines is defined as BPX shape length.  }
    \label{fig:profile}
\end{figure*}

 The top panel of Figure \ref{fig:BSPSBU} shows the time evolution of bar strength. The bar strength continuously increases with time during the initial growth of the bar until 1.43 Gyr. The bar strength shows an abrupt decrease until 1.45 Gyr and it starts increasing continuously again until 2.2 Gyr. The bar strength does not grow between 2.2 to 3.2 Gyr though shows a short plateau distribution in this interval. Then beyond 3.2 Gyr bar strength grows and saturates in the secular evolution of the disk. Figure \ref{fig:barlength} shows the growth of bar length with time. The bar length decreases between 1.37 to 1.54 Gyr and shows a plateau between 2.2 to 3.2 Gyr like the bar strength evolution with time. The bar length continuously increases with time in secular evolution, unlike bar strength which saturates. We show the face-on and edge-on maps of bar evolution throughout the simulations in Figure \ref{fig:Time_evolution_face_on_and_edge_on} and discuss in Appendix \ref{Time_Evolution}.  The second panel of Figure \ref{fig:BSPSBU} shows the time evolution of the pattern speed of the bar after the bar has formed ($A_2/A_0 >0.35$). We find that pattern speed decreases with time after the bar gains strength by transferring angular momentum to the outer disk and dark matter halo \citep{Kataria.Shen.2022}. The pattern speed shows a sharp decrease at 1.45 Gyr coinciding with the time of the sudden decrease in bar strength. In the secular phase of bar evolution, pattern speed decreases smoothly until the end of the simulation. The third panel of Figure \ref{fig:BSPSBU} shows the time evolution of the vertical buckling strength which shows two clear peaks at 1.45 Gyr and 2.68 Gyr followed by small oscillations in secular evolution. These two buckling events, differing in 30 \% in their strength, are represented by dashed and dotted vertical lines respectively which have the above-mentioned imprints on the bar strength and pattern speed evolution. The last panel of Figure \ref{fig:BSPSBU} shows the time evolution of $\sigma_z/\sigma_r$ which is calculated within $2R_d$ region of the disk. The  $\sigma_z/\sigma_r$ shows a decrease as the bar grows in the disk and results in the increased radial dispersion of the disk. Afterwards, there is an abrupt and huge increase in the $\sigma_z/\sigma_r$ value ($\approx$ 0.65 to 0.77) during the first buckling at 1.45 Gyr. Further, $\sigma_z/\sigma_r$ starts decreasing until the second buckling. In the second buckling event at 2.68 Gyr, there is a slow and relatively smaller increase in $\sigma_z/\sigma_r$ values ($\approx$ 0.68 to 0.72). In the secular phase of bar evolution, $\sigma_z/\sigma_r$ decreases and saturates at the end of the simulation.

\subsection{Buckling events, corresponding BPX shape lengths and kinematic signatures}
Figure \ref{fig:Edgeon_maps} animates the edge-on stellar mass density evolution of the disk during the two buckling events. We quantify the buckling time scales of individual events by calculating the time duration in which $A_{Buckle}>0.06$. The first buckling lasts for a shorter time around 220 Myr ($\approx$ 1.32 to 1.54 Gyr) while the second buckling lasts for a longer time around 1 Gyr ($\approx$ 2.0 to 3.01 Gyr). The left panel of Figure \ref{fig:Edgeon_maps} shows the bending of the bar during the first buckling in the edge on view. This buckling event is confined within the 3 kpc region of the disk, resulting in a BPX of similar size. On the other hand, the right panel of Figure \ref{fig:Edgeon_maps} shows that second buckling spans over a larger radius of around 6 kpc which results in the larger and more pronounced BPX shapes. We quantify the full length of BPX by measuring the difference in peaks of the line profiles along $z=$ 1.5 kpc or -1.5 kpc in the x-z plane. Figure \ref{fig:profile} shows the line profiles along $z=$ 0,1.5,-1.5 kpc lines in the x-z plane. We find that the first buckling results in a full  BPX shape length of around 5.8 kpc which is around two-thirds of the full bar length ($\approx$ 9 kpc) at this time.  On the other hand the second buckling results in full BPX shape length of around 8.6 kpc which is also a two-thirds of full bar length ($\approx$13 kpc). 

\begin{figure*}
    \centering
    \includegraphics[scale=0.4]{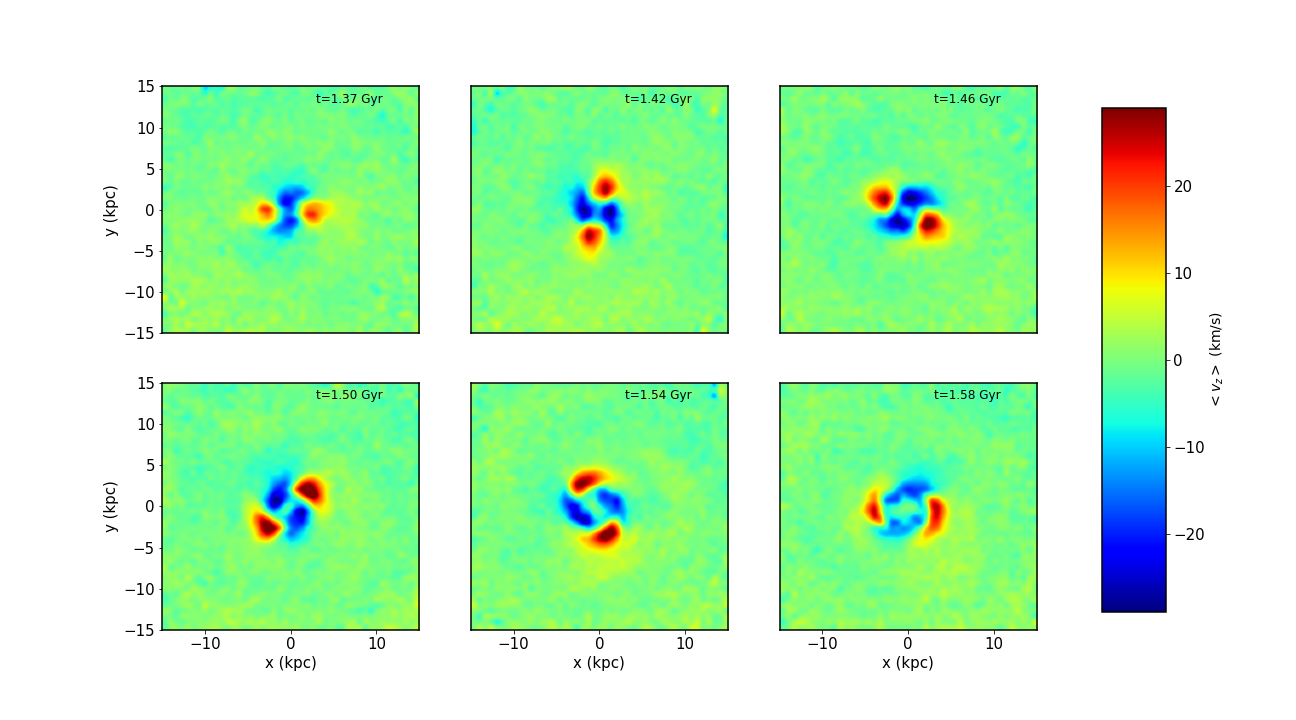}
    \caption{Time evolution of mean $V_z$ maps during first buckling event when disk is seen face-on.}
    \label{fig:First_buckling_face_on_vz_maps}
\end{figure*}

Figure \ref{fig:first_buckling_face_on} shows the face-on map of stellar density evolution during the first buckling event. We find that the bar length decreases from 1.37 Gyr to 1.54 Gyr which is quite clear from isodensity contours plotted in dashed lines, resonating with the direct bar length measurements in Figure \ref{fig:barlength}. We find that during this buckling event, the spiral modes are enhanced from the edge of the bar which is evident from isodensity contours at 1.42 Gyr. We have also plotted the mean $V_z$ map of star particles along the face-on view of the disk during the first buckling event in Figure \ref{fig:First_buckling_face_on_vz_maps}. This kinematic map is motivated by the recent detection of quadrupolar patterns emerging out in the line-of-sight velocities of almost face-on galaxies in the MaNGA survey \citep{Xiang.etal.2021}. We observe the quadrupolar patterns in mean $V_z$ maps which originate from the very central region at the beginning of the buckling event and expand out at the end of buckling event. The peak strength of quadrupolar patterns in mean $V_z$ coincides with the peak of buckling amplitude ($A_{Buckle}$). We can notice that these quadrupolar patterns last for the whole buckling event. Finally, these patterns for the first buckling are compact, symmetric, and more pronounced compared to the mean $V_z$ of disk background.

\begin{figure*}
    \centering
    \includegraphics[scale=0.4]{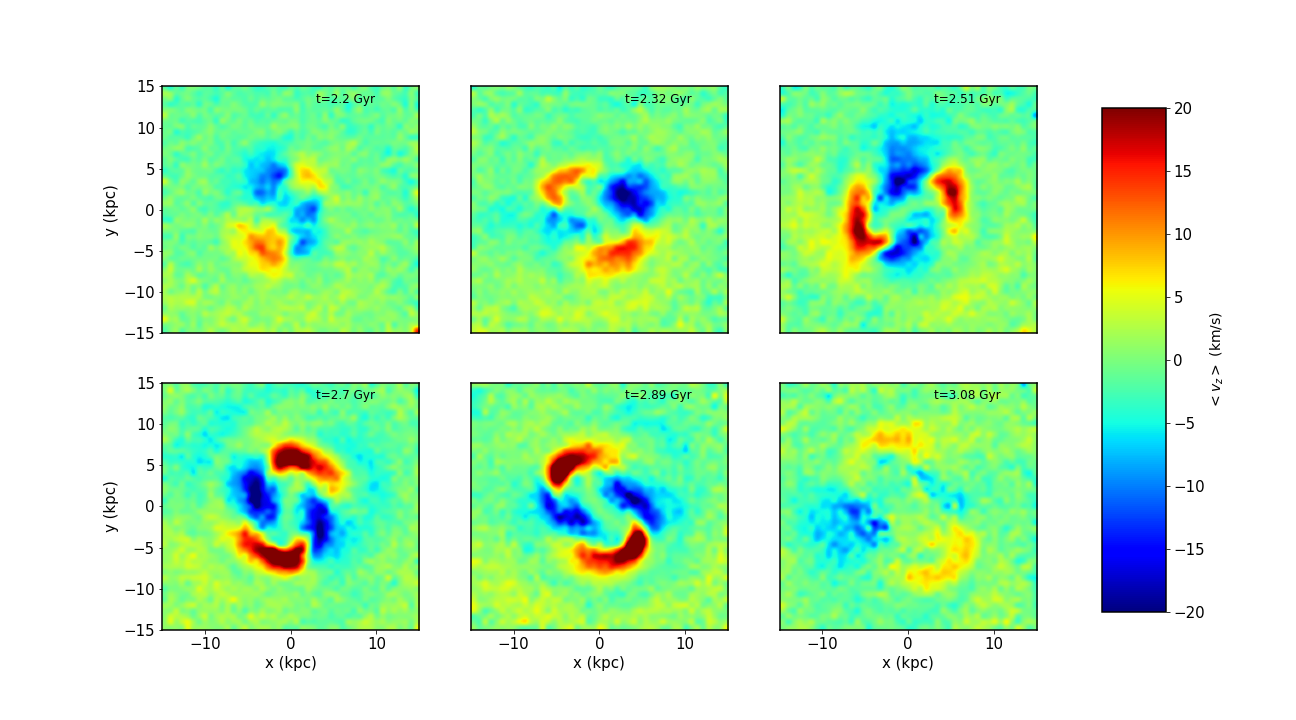}
    \caption{Time evolution of mean $V_z$ map during second buckling event when disk is seen face-on.}
    \label{fig:second_Buckling_Face_on_vz}
\end{figure*}

\indent Similarly, we examine the face-on stellar density with isodensity contours evolution and the quadrupolar patterns in the mean $V_z$ maps evolution during the second buckling in Figures \ref{fig:second_Buckling_Face_on} and \ref{fig:second_Buckling_Face_on_vz} respectively. We find that the semi-major axis of the bar remains the same ($\approx$ 6.2 kpc) during the whole secondary buckling event. Isodensity contours in Figure \ref{fig:second_Buckling_Face_on} show the weak spiral modes in the outer region of the disk which are not connected to the edge of the bar ($\approx$ 10 kpc). The quadrupolar patterns of the mean $V_z$ show a weak signal at the beginning of the buckling which becomes strengthened during the peak of buckling and fades away at around 3.08 Gyr. The region encompassing the quadrupolar patterns increases with radius from the beginning to the end of buckling event though remains confined within the semi-major axis of bar. In comparison to the first buckling, these quadrupolar patterns during the second buckling are extended, non-symmetric, and less pronounced compared to mean $V_z$ of the disk background. 

\subsection{Time Evolution of Fourier modes ($m=$ 2 and 4) throughout the disk } 
\begin{figure*}
    \centering
    \includegraphics[scale=0.4]{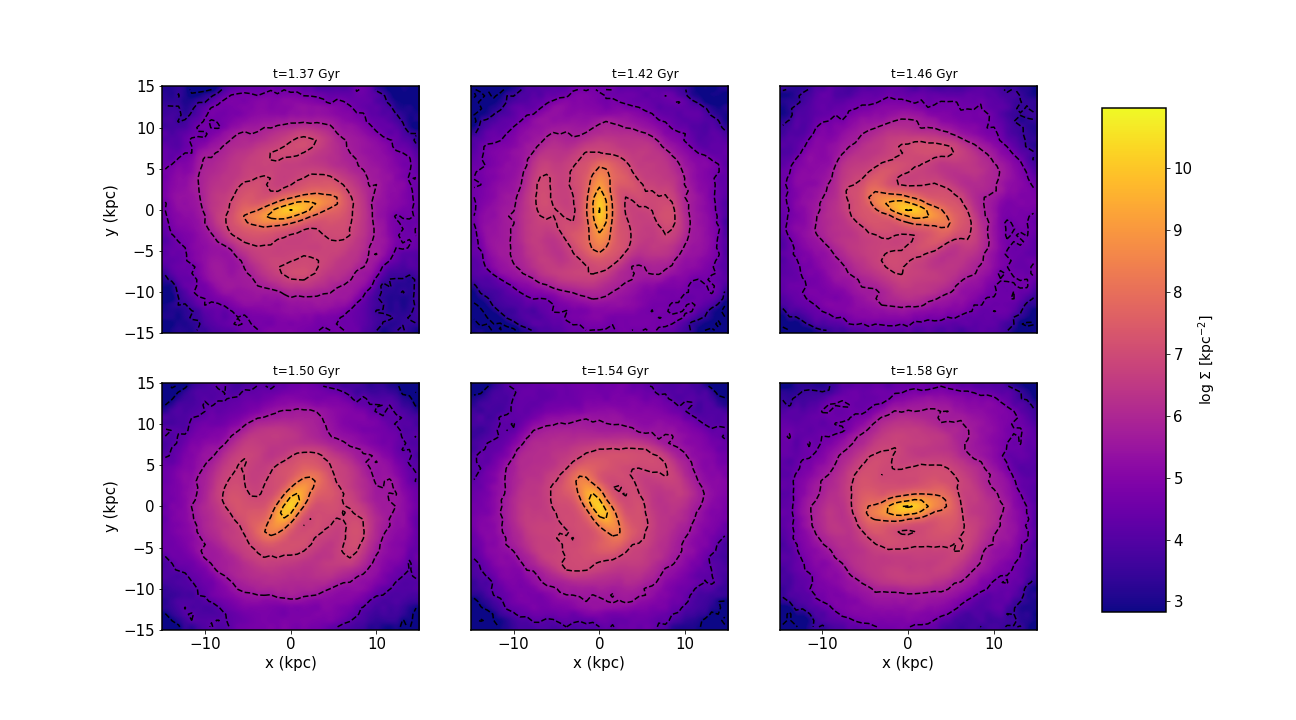}
    \caption{Time evolution of face-on stellar mass density maps around the first buckling event. The dashed lines represent the isodensity contours.}
    \label{fig:first_buckling_face_on}
\end{figure*}

\begin{figure*}
    \centering
    \includegraphics[scale=0.4]{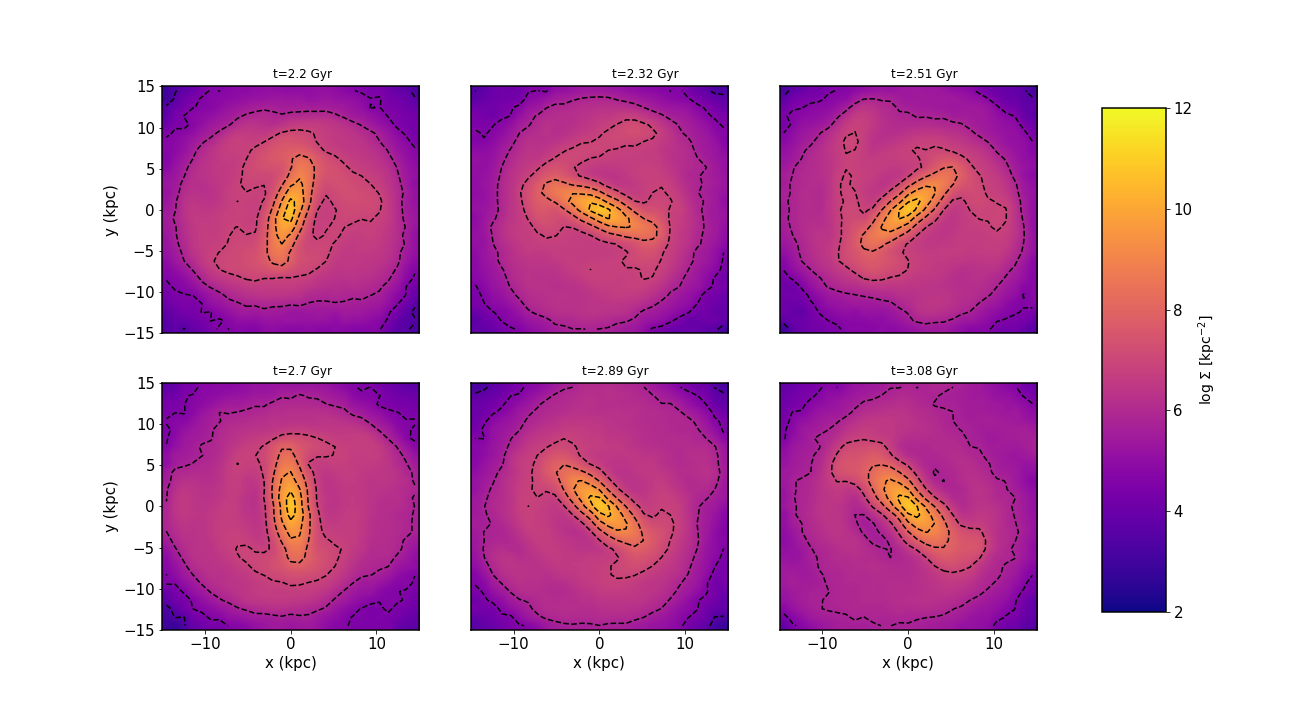}
    \caption{Time evolution of face-on stellar density maps during the second buckling. The dashed lines correspond to isodensity contours.}
    \label{fig:second_Buckling_Face_on}
\end{figure*}

\begin{figure*}
   \begin{tabular}{c|c}      \subfigure{\includegraphics[width=0.5\textwidth]{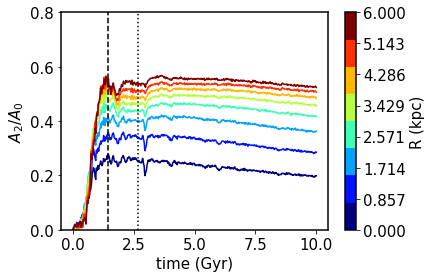}} 
   \subfigure{\includegraphics[width=0.5\textwidth]{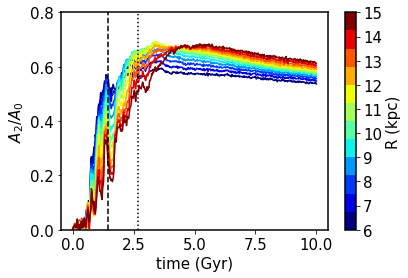}} 

   \end{tabular}

    \caption{Left panel shows the time evolution of m=2 Fourier mode ($A_2/A_0$) with time in concentric rings of increasing radius until 6 kpc of disk radius. The right panel show the same m=2 mode for concentric rings at a radius ranging from 6 kpc to 15 kpc. Here vertical dashed line corresponds to the first buckling ($\approx$ 1.45 Gyr) and the vertical dotted line corresponds to the second buckling event ($\approx$ 2.68 Gyr).}
    \label{fig:A2byA0}
\end{figure*}

\begin{figure*}
   \begin{tabular}{c|c}
    
    \subfigure{\includegraphics[width=0.5\textwidth]{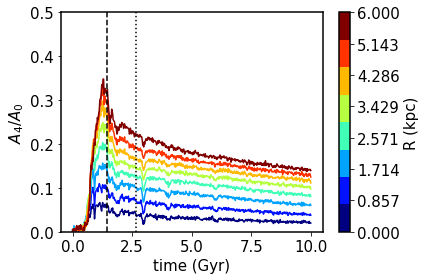}}
       
    \subfigure{\includegraphics[width=0.5\textwidth]{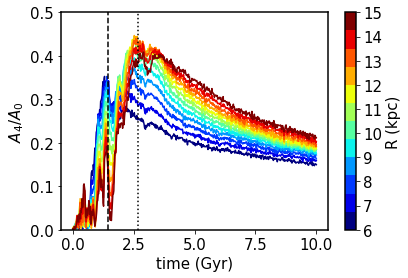}} 

   \end{tabular}

    \caption{ This figure is similar to Figure \ref{fig:A2byA0} though capture $m=4$ Fourier mode in the disk. }
    \label{fig:A4byA0}
\end{figure*}

We observe that two buckling events have pronounced effects on the BPX shapes of bars as seen in Figure \ref{fig:profile}. In order to understand the impact of buckling throughout the disk we plot the $m=$ 2 and 4 modes in concentric rings of width equal to  0.75 kpc throughout the disk in Figures \ref{fig:A2byA0} and \ref{fig:A4byA0} respectively. The left panel of Figure \ref{fig:A2byA0} shows that the first buckling does not affect the growth of $A_2/A_0$ mode in the inner disk region (radius $<$ 6 kpc) while the second buckling shows a sudden dip in $A_2/A_0$ mode in this region. The division of the inner and outer regions is motivated by the difference in the time evolution of the overall pattern of $A_2/A_0$ and the dividing radius coincides with the typical bar semi-major axis length during the second buckling event. The right panel of Figure \ref{fig:A2byA0} shows that two buckling events have reverse trend,  for a sudden dip in the time evolution of $A_2/A_0$ mode, in the outer disk (radius $>$ 6 kpc) compared to the inner part of the disk. This is interesting to observe that the first buckling which is confined to the central region of the disk (radius $<$ 3 kpc) affects the $m=2$ mode at the outer region of the disk. This is correlated with the growth and decay of spiral arms at the edge of bar (bar semi-major axis $\approx$ 4 kpc) during the first buckling as seen in Figure \ref{fig:first_buckling_face_on}. On the other hand, the second buckling which spreads across the larger regions of the disk (radius $\approx$ 5 kpc) affects the $m=$ 2 mode of the central region of the disk. This is correlated with the inner region of the bar showing peanut features during the end of this buckling event as seen in Figure \ref{fig:second_Buckling_Face_on} at 3.08 Gyr. The time span of buckling events correlates inversely with the length scale of the disk where $m=$ 2 mode shows a sudden dip during its time evolution. For example the shorter time span of the first buckling shows a dip in the $m=$ 2 mode at large scale lengths of the disk and vice-versa. These modes show particular isophotal features in the face-on density maps which depend on the sequence or time span of buckling.

Figure \ref{fig:A4byA0} shows that the dip in $m=$ 4 mode during first buckling increases with radius though present throughout the disk. The second buckling results in much smaller dips in $m=4$ Fourier modes (compared to the first buckling) which further reduces with disk radius except the outermost radius ($\approx$ 15 kpc). This suggests that the box shape of the face-on disk is mostly affected by the first buckling mostly at the outer regions of the disk which can be gauged in isodensity contours of Figure \ref{fig:first_buckling_face_on}. Both the first and second buckling affect the inner region box shape of a face-on disk which can be readily noticed in Figures \ref{fig:first_buckling_face_on} and \ref{fig:second_Buckling_Face_on}.  

\subsection{Time Evolution of $\sigma_x$ and $\sigma_z$ with radius along the bar region }
\begin{figure*}
   \begin{tabular}{c|c}
    
    \subfigure{\includegraphics[width=0.5\textwidth]{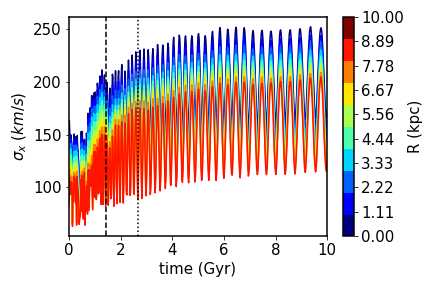}}
       
    \subfigure{\includegraphics[width=0.5\textwidth]{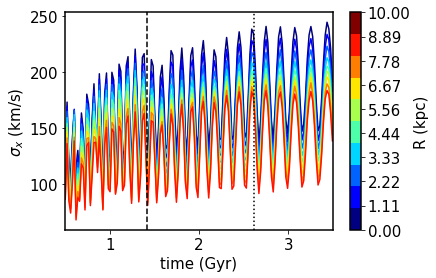}} 

   \end{tabular}

    \caption{Time evolution of $\sigma_x$ within bins of 1 kpc radius along the bar major axis which covers $|x|$< 10 kpc,$|y|$< 3 kpc, $|z|$< 4 kpc. Right panel is the high resolution time evolution up to 3.5 Gyr. The dashed and dotted line corresponds to first and second buckling events respectively.}
    \label{fig:sigmax}
\end{figure*}

Left panel of Figure \ref{fig:sigmax} shows the time evolution of $\sigma_x$ within the bins of size 1 kpc along the bar major axis covering a box of $|x|$< 10 kpc, $|y|$< 3 kpc, $|z|$< 4 kpc. The time evolution of $\sigma_x$  at any radius shows fluctuations which is due to the rotation of the bar. The frequency of this fluctuation decreases as the bar slows with time. We find that as the bar grows until 1.4 Gyr, the peak value of $\sigma_x$ increases and shows a sudden dip at the time of first buckling ($\approx$ 1.45 Gyr) at all radii which is clear in high-resolution plot in the right panel of Figure \ref{fig:sigmax}. Further, the peak value of $\sigma_x$ increases and shows a short plateau during the second buckling around 2.68 Gyr. Finally, $\sigma_x$ increases smoothly in the secular phase of bar evolution. 

\begin{figure*}
   \begin{tabular}{c|c}
    
    \subfigure{\includegraphics[width=0.5\textwidth]{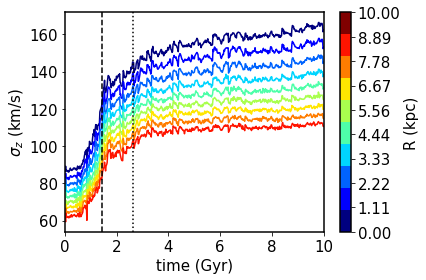}}
       
    \subfigure{\includegraphics[width=0.5\textwidth]{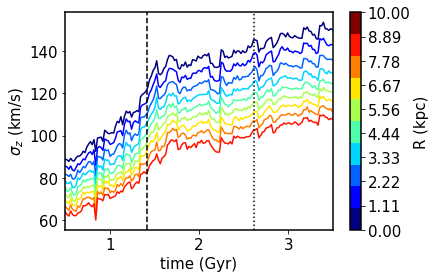}} 

   \end{tabular}

    \caption{Time evolution of $\sigma_z$ within bins of 1 kpc radius along the bar major axis which covers $|x|$< 10 kpc,$|y|$< 3 kpc, $|z|$< 4 kpc. The right panel is the high resolution time evolution up to 3.5 Gyr. The dashed and dotted line corresponds to first and second buckling events respectively.}
    \label{fig:sigmaz}
\end{figure*}

The left panel of Figure \ref{fig:sigmaz} shows that $\sigma_z$ increases slowly at all radii as the bar grows and shows a sudden rise as the bar buckles at 1.45 Gyr. Afterwards, $\sigma_z$ increases at a slower rate during the second buckling around 2.68 Gyr and the rate of increase further reduces in the secular evolution phase of the bar at all radii. The rate of increase in $\sigma_z$ is noticed clearly in the high-resolution plot in the right panel of Figure \ref{fig:sigmaz}.  This shows that the first buckling is efficient in transporting the planar energy to the vertical plane compared to the second one.

\section{Discussion} \label{discussion}

Observations show that only a handful of galaxies (only 8 galaxies) have been caught in ongoing buckling action \citep{Erwin.Debattista.2016, Li_2017,Xiang.etal.2021}. \cite{Erwin.Debattista.2016} report two ongoing buckling events in their sample of 44 galaxies using photometric and kinematic signatures which suggest the buckling fraction to be around 4.5 \%. Their toy model, which correctly predicts the frequency of buckling events with their observations, assumes the buckling timescale to be wide-ranged between 0.5 to 1 Gyr. Further, this study is limited by the sample size for a robust accounting of the frequency of ongoing buckling and bound on the timescale of buckling events. In a recent advancement that involves a novel approach, \cite{Xiang.etal.2021} report 5 ongoing buckling in a sample of 434 galaxies by detecting quadrupolar patterns in line-of-sight (LOS) velocities in the MaNGA survey having integral field spectroscopic data (IFU). This study also conveys the message of missing detection of buckling events in their sample due to limits on signal-to-noise ratios. Their toy model, based on a semi-analytical bar evolution model, predicts a lower bound on the buckling timescale to be 130 Myr. These semi-analytical calculations do not account for the secondary buckling events which can affect the detection probability and hence the lower bound on the buckling timescales. In our N-body model, the bar undergoes two buckling episodes lasting for around 220 Myr and 1 Gyr over a time span of 9 Gyr of bar evolution. Therefore, the detection probability of buckling events must be higher than the recently observed probability of these events, given the bar can last for buckling around an order of magnitude larger than predicted by \cite{Xiang.etal.2021}.

The secondary buckling events are less common in the galaxies with massive classical bulges \citep{Smirnov.2019} and high gas fraction \cite{Lokas.2020}. This certainly favours the toy models by \cite{Xiang.etal.2021} which does not account for the second buckling event. Although disks with less massive classical bulges can still undergo multiple buckling events which needs further explorations. Further, the strength of tidal interactions enhance the strength \citep{Lokas.2019A} and timescale \citep{Lokas.2019A,Kumar.etal.2021} of buckling which has not been accounted for the toy model of \cite{Xiang.etal.2021}. This implies the enhanced detection probability of buckling events given the tidal interactions has been prominent in the late universe \citep{Sinha.Bockelmann.2012}. Further, \cite{KumarA.et.al.2022} report three buckling events in a single galaxy evolution with prolate dark matter, for which the time of all the buckling events is a significant factor of the bar evolution time. This certainly implies a higher detection probability of such events which certainly need to be accounted in toy model of \cite{Xiang.etal.2021}.   

The time span of buckling and a number of successive buckling events are the key parameters for spotting the ongoing buckling action of bars in observed galaxies. In this work, we have looked into the role of successive buckling events on the potential observable properties of bar and disk.  The kinematic observable signature involves the presence of quadrupolar patterns on the LOS velocity in the near face-on projection of disk. The longevity of these signatures is directly proportional to the time of buckling as evident from Figures \ref{fig:First_buckling_face_on_vz_maps} and \ref{fig:second_Buckling_Face_on_vz}. Further, the shorter time span of buckling affects $m=$ 2 and $m=$ 4 Fourier modes at larger scale lengths of disk. The observable signatures involve the growth and decay of spiral arms at the edge of bar ($m$=2) as well as the pronounced box shape of disk ($m$=4) during the first buckling event as seen in Figure \ref{fig:first_buckling_face_on}. This can provide secondary signatures on the time spans of the buckling or successive buckling events in the photometric observations.

The bending of the bar has mostly been understood in terms of firehose instability \citep{Raha.etal.1991, Binney&Tremaine_2008}. In a very recent study, \cite{Xingchen.et.al.2023B} propose the resonant action due to a certain common fraction of vertical (2:1) and planer (2:1) resonances of the disk leading to vertical buckling of the bar. In this context, it would be interesting to analyse the allowed ranges of buckling time spans with both firehose instability and resonant action phenomenon. This can certainly shed light on the origin of buckling instability.      

\cite{Collier.2020} discussed the role of violent buckling compared to non-violent one where violent buckling has a lower timescale compared to non-violent one. The violent buckling does not destroy the bar and washes out quickly without affecting the bar in secular evolution while the non-violent one destroys the bar by sustaining bending modes for longer times.  The current work has only violent buckling events ($A_{Buckle}>0.05$) given that do not destroy the bar irrespective of the time span of the bending mode. This highlights that the buckling mode lasting longer in the bar is not sufficient to destroy the bar.  

\section{Conclusions} \label{conclusion}
In this article, we perform an N-body simulation of a Milky Way-type disk galaxy having a stellar disk and dark matter halo components. We report that the disk forms a bar within a Gyr of evolution and undergoes two successive vertical buckling events occurring at 1.45 Gyr and 2.68 Gyr respectively. The first buckling and second buckling have time spans of around 220 Myr and 1 Gyr respectively though the second buckling is slightly stronger than the first one around $30\%$.  We have looked for the role of these two successive buckling events on the observable properties of the bar and disk. The main results are listed below:

\begin{itemize}
    \item First buckling shows an abrupt change in the time evolution of the pattern speed while second buckling does not show this. This abrupt change in pattern speed leads to a faster decrease in pattern speed followed by sudden increase during the buckling event.
    \item First buckling with a short time span leads to a shorter BPX bulge ($\approx 5.8$ kpc) and the second buckling with a larger time span leads to a longer BPX bulge ($\approx$ 8.6 kpc).  BPX bulge full lengths corresponding to successive buckling events are around two-thirds of the full bar lengths at the end of individual buckling events.

    \item  The kinematic signature for the bar buckling is observed in the LOS mean velocity for a nearly face-on galaxy disk. Both the buckling events show the quadrupolar patterns in mean LOS velocity which last thoughout the buckling event. The quadrupolar patterns in the first buckling are compact, symmetric, and pronounced compared to those in the second buckling, which are extended, non-symmetric, and less pronounced.
    \item The Fourier mode ($m$=2) in the x-y plane decreases sharply for the outer regions of the disk ($r>6$ kpc) during the first buckling while the bar bends mostly in the inner region ($r<3$ kpc) region. This is correlated with the growing and decaying spiral features at the end of the bar, an observable feature in the isodensity contours. For the second buckling, the Fourier mode ($m$=2) shows a sharp dip in the inner regions of the disk ($r<4$ kpc) while the bar bends until $r\approx 5$ kpc. This is linked with the face-on butterfly shape of the bar in the central region of the disk during the second buckling event. 

    \item The dip of the Fourier mode ($m$=4) during the first buckling event increases with radius, suggesting the outer part of the disk loses face-on boxy features more than the inner bar of the disk. In comparison, the second buckling leads to a much smaller dip in the Fourier mode ($m$=4) which decreases with radius though the mode recovers just after the second buckling.

    \item First buckling leads to a sharp dip in radial dispersion along the bar and a sharp rise in vertical dispersion throughout the disk, suggesting very rapid transfer of energy from a horizontal plane to the vertical plane of the disk. The second buckling leads to the flattening of radial dispersion along the bar and a slow increase in vertical velocity dispersion throughout the disk, suggesting a relatively slower transfer of energy across horizontal and vertical planes.   
\end{itemize}

Given the combination of photometric and kinematic observable signatures, these methods are insightful for existing IFU surveys from MaNGA data \citep{Xiang.etal.2021} to detect the ongoing bar buckling.  The proposed methods empower ongoing and upcoming observations for gauging whether the observed ongoing buckling event is a successive buckling event or not. Further, these methods can be easily applied to the high redshift IFU capability of JWST \citep{Boker.et.al.2022}, for the detection of ongoing buckling events.

In connection with the recent findings about resonant action being the cause of buckling \citep{Xingchen.et.al.2023B} in comparison to the well-known notion of firehose instability \citep{Raha.etal.1991}, we will be discussing this comparison elsewhere in the context of the time span of buckling.

\section*{Acknowledgements}
SKK thanks anonymous referee for the insightful comments to improve the content in the draft. The research presented here is partially supported by the National Key R\&D Program of China under grant No. 2018YFA0404501; by the National Natural Science Foundation of China under grant Nos. 12025302, 11773052, 11761131016; by the ``111'' Project of the Ministry of Education of China under grant No. B20019; and by the China Manned Space Project under grant No. CMS-CSST-2021-B03. SKK thanks Volker Springel for the Gadget code which we used to run our simulations. SKK thanks Juntai Shen, Zhi Li, Rui Guo, Xingchen Li and Yirui Zheng for the scientific discussion which helped to progress in this work.  This work made use of the Gravity Supercomputer at the Department of Astronomy, Shanghai Jiao Tong University, and the facilities of the Center for High Performance Computing at Shanghai Astronomical Observatory.
software used: numpy \citep{Harris.et.al.2020}, matplotlib \citep{Hunter.2007}, pynbody \citep{Pynbody.2013} and astropy \citep{Astrop.collaboration.2018} packages. 

\section*{Data Availability}

The data will be shared on a reasonable request to the author.



\bibliographystyle{mnras}
\bibliography{example} 

\appendix 
\section{Time Evolution of Face-on and Edge-on Bar} \label{Time_Evolution}
\begin{figure*}
    \centering
    \includegraphics[scale=0.50]{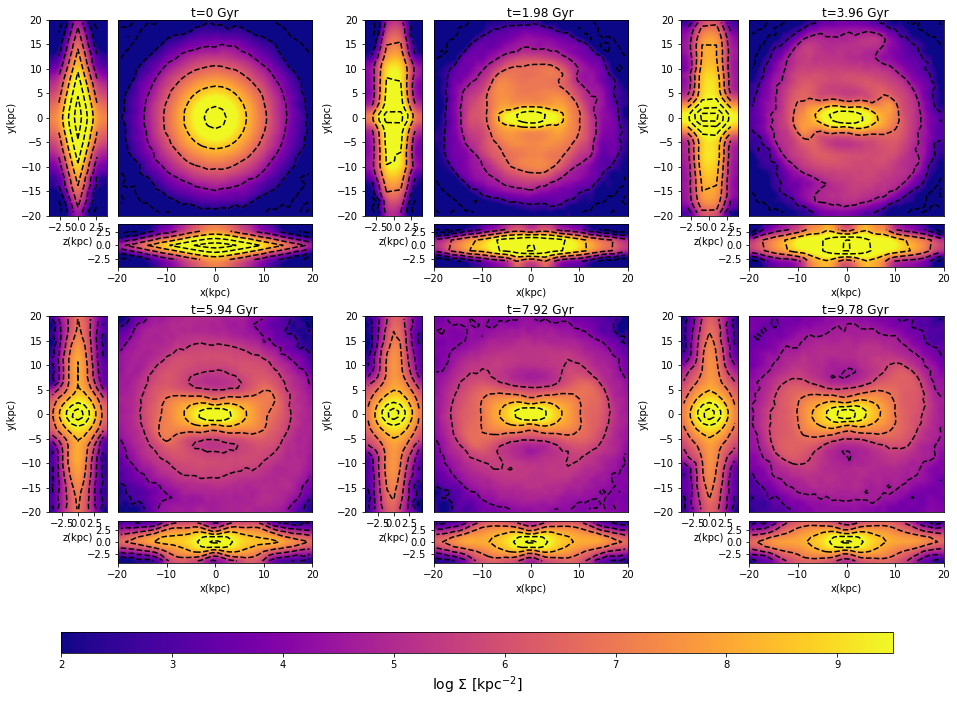}
    \caption{Time evolution of the face-on and edge-on stellar density maps throughout the simulation. Here bar is oriented manually along the horizontal axis. The dashed lines represent the isodensity contours.}
    \label{fig:Time_evolution_face_on_and_edge_on}
\end{figure*}

Figure \ref{fig:Time_evolution_face_on_and_edge_on} shows the time evolution of the face-on and edge-on shapes of bar throughout the simulation up to 9.78 Gyr. We can see that bar length grows with time which is also quantified in Figure \ref{fig:barlength}. In the face-on map, we can see that the bar thickens along the minor axis and shows peanut-pronounced butterfly shapes in the central region which is evident from the isodensity contours. In the edge-on shapes of the bar, we can also notice the growth till the end of simulations.




\bsp	
\label{lastpage}
\end{document}